\documentclass[11pt, oneside]{article}       % use "amsart" instead of "article" for AMSLaTeX format
\usepackage{geometry}                        % See geometry.pdf to learn the layout options. There are lots.
\usepackage{graphicx}                % Use pdf, png, jpg, or eps§ with pdflatex; use eps in DVI mode
\usepackage{amssymb}
\usepackage[super, numbers]{natbib}

\newcommand{\R}{\mathcal{R}}

\title{Skeptical Notes on a Physics of Passage}
\author{Nick Huggett$^a$\\\ \\$^a$University of Illinois at Chicago}
\date{}                            % Activate to display a given date or no date

\begin{document}
\maketitle

\noindent CONTACT:

Department of Philosophy, MC 627

601 S.Morgan St, Room 1421

University of Illinois at Chicago

Chicago, IL 60607\\

huggett@uic.edu

312-996-3022

312-413-2093\\


\noindent ABSTRACT: This paper investigates the mathematical representation of time in physics. In existing theories time is represented by the real numbers, hence their formal properties represent properties of time: these are surveyed. The central question of the paper is whether the existing representation of time is adequate, or whether it can or should be supplemented: especially, do we need a physics incorporating some kind of `dynamical passage' of time? The paper argues that the existing mathematical framework is resistant to such changes, and might have to be rejected by anyone seeking a physics of passage. Then it rebuts two common arguments for incorporating passage into physics, especially the claim that it is an element of experience. Finally the paper investigates whether, as has been claimed, `causal set theory' provides a physics of passage.\\

The papers in this volume are based on talks given at a meeting Cape Town, in response to the question, `do we need a physics of passage?' First, let's take it that the question asks whether we need a \emph{new} physics, incorporating passage in a way that existing physics does not -- especially, does a quantum theory of gravity require passage? Second, we are asking about a physics of passage, so the question is about the possible need for a \emph{mathematical representation} of passage.

`Passage' does not have an unequivocal meaning, so there is a different question for each meaning; however, my answer will be largely negative for each of the senses that I canvass, though for varying reasons. Central to the discussion will be careful attention to the ways in which time is represented in existing physics, in the first place by the `minimal model', which represents time by the real numbers. Some senses (or aspects) of passage are already represented in that model (or in existing extensions of it): so they don't require a new physics of passage. However, `passage' often means something stronger (though exactly what is hard to pin down), so I will also show -- with one possible exception -- how some prominent arguments for the inadequacy of the existing physics of passage fail. Moreover, I want to make clear that attempts to develop a new physics of time typically require radical revisions in the most basic mathematical assumptions of physics: most paths to passage demand a willingness to start almost from scratch.

Though I have a negative thesis, presenting that conclusion is the secondary aim of the paper. Rather, my main purpose is to explicate some philosophical conceptions of passage in a form that I hope will be useful to physicists interested in a physics of passage: though my survey of conceptions of passage will not be exhaustive. History teaches us that major developments in physics -- Newtonian gravity, statistical physics, relativity, quantum mechanics, and so on -- require major conceptual revisions, and hence involve philosophical thought. As a small contribution to such a project I offer some distinctions and clarifications of our existing conception of time (and in one case of a new conception).

As I said, the notion of `passage' is equivocal. For instance, in arguing for passage, Ellis sometimes \cite{ellis2013arrow} (but not always) takes his target to be Barbour's view that there is only one instant. But while I deny passage, I do not deny that there is more than one moment of time: in tensed language, there were past times and there will be future times. (Of course, they don't all happen at the same time!) If `passage' just means that there is more than one instant, then any physics with timelike extension is \textit{ipso facto} a physics of passage. But then we don't \textit{need} a physics of passage, as our existing spacetime physics already accommodates this conception. Nor does temporal extension seem to fully capture the concept of passage, though it is one aspect: in addition, passage denotes, for instance, change and asymmetry. But isn't change just a matter of a system being in different states at different times? And that is already a feature of physics (with a suitable relativistic gloss if necessary). Similarly, the theory of weak interactions already has a temporal asymmetry. 

So clearly what is at stake is some even thicker notion of passage. In Cape Town, Price gave a talk based on a paper \cite{price2009flow} in which he distinguished three notions of temporal passage (or `flow of time'): the third is that passage is `dynamic' in some robust sense. Even proponents of this kind of passage often agree with Price that the conception cannot be completely described, simply `pointed at' (like pain, perhaps), so I won't try either. (Below I will suggest that alleged experiences of dynamical passage, while real experiences, are not of passage at all.) However, our mathematical representations of time are -- as uninterpreted mathematical models -- atemporal, unchanging: some commentators have the intuition that they therefore fail to represent the dynamical aspect of passage. Do they? And if so, could we change our mathematics to make it intrinsically `temporal'? Those are my questions. 

The paper develops the considerations just made, in terms of the mathematical role of the various conceptions of passage, to make clear the relevance for physical theory. \S\ref{MMT} presents a stripped-down mathematical representation of time, the `minimal model', and shows both how it incorporates important aspects of passage, and how it strongly resists modifications aimed to represent a thicker, `dynamic' sense of passage. \S\ref{EMMT} further discusses proposals for extending the minimal model. Some, such as a temporal asymmetry, can be readily incorporated; others, connected to the alleged dynamical conception, should not or cannot be incorporated. Finally, in \S\ref{CST} I use the ideas of the paper to discuss what it would take for causal set theory to implement a notion of passage.

\section{The Minimal Model of Time}\label{MMT}

According to the minimal model of time, time is adequately represented by the real numbers. (In fact the model is not entirely adequate, as we shall discuss later, but the reals do capture the core aspects with which this section is concerned. Especially, in this section it will be convenient to focus on non-relativistic spacetimes, in which the reals parameterize an absolute time.) Three points about the model (or its easy extensions): first, the minimal model is the basic representation of time assumed in all standard physics, classical and quantum; second, all concrete features that we might associate with temporal passage can be accounted for by the model (hence are compatible with standard physical treatments of time); third, the model, if not excluding the thicker sense of passage alluded to above, is inhospitable to it. The first point seems straight-forward. However, it is perhaps so straight-forward as to be overlooked and its significance under-estimated; this section will describe some familiar mathematical facts to make explicit the theory of time they embody, and start to address the second two points.\\

A mathematical theory of passage ought to include multiple times somehow, for there can be no representation of passage in a model of a single instant. If, say, only noon on Tuesday is represented in any way, then there can be no representation of time passing from one time to \emph{another}.$^a$ This feature is well represented by the reals, since the different points of $\R$ do represent different moments. In this regard even discrete time represents time passing. But there is more to passage than multiple times: for instance, passage should allow for change. Of course it does in the minimal model. Formally, the state of a system is represented by some function of the reals taking on some value: for instance, coordinates that represent a position. For simplicity, we'll consider a (continuous) function whose values are real -- $f:\R\to\R$ -- but that assumption is not significant in this discussion. Physically then, $f(t)$ represents the state a system takes at a given moment, labelled with $t\in\R$. Change in the system is simply represented by the function taking on different values at different times. Again, this famous `at-at' theory of change works equally well for discrete time and $\R$-time. However, in the latter case a new aspect of passage must be represented.

The question is to understand how some quantity `reaches' a given value as time passes from an arbitrary initial time to the present. This is an ancient problem of course, for instance raised in many of Zeno's paradoxes of motion: in general, it shows one way in which change and passage are linked. Formally we would write $f(t)\to f(t_{now})\ \mathrm{as}\ t\to t_{now}$ to express this state of affairs: for the point I am making, we need to leave open the interpretation of such an expression, and so do \emph{not} read it in the way familiar from elementary calculus. Instead, the question is of the significance of $t\to t_{now}$, which naively appears to represent the passage of time. If one follows this line of thought, then understanding how change is formally possible requires assigning a precise mathematical meaning to $t\to t_{now}$, and hence giving a mathematical representation to passage. (Note that this question does not arise for discrete time according to the at-at theory. There is no additional formal question of how a function reaches a certain value at a certain time: it's simply the case that at the given times it is at the appropriate values. To ask for more is just to say that the standard account of discrete functions is inadequate.)

But of course no such independent meaning was ever given. Instead Cauchy (and his precursors) gave us the famous $(\epsilon,\delta)$-definition of $f(x)\to a\ \mathrm{as}\ x\to \mathrm{x}$: $f(x)\to a\ \mathrm{as}\ x\to\ \mathrm{x}$ if for any $\epsilon>0$ there is some $\delta>0$ such that $|f(x)-a|<\epsilon$ for all $x$ satisfying $|\mathrm{x}-x|<\delta$. In other words, to say that $a$ is the limit at x means that however close one wants to get to $a$, as long as $x$ is close enough to x, then $f(x)$ will be as close as one desired.

Instead of providing a logically independent meaning for $x\to$ x, this definition relates function and independent variable in a way that cannot be split up as envisioned in the previous line of thought. Courant and Robbins make the familiar point very clearly: ``The independent variable does not move, it does not `tend to' or `approach' a limit $a$ in any physical sense \dots. No part of [the $(e, \delta)$ definition], e.g. `$x \to$ x' has a meaning by itself.'' Or, to paraphrase in terms more appropriate to the present discussion, ``Time does not move, it does not `tend to' or `approach' the present in any physical sense. \dots. No part of the $(\epsilon,\delta)$-definition of $f(t)\to f(t_{now})\ \mathrm{as}\ t\to t_{now} $, e.g. `$t \to t_{now}$' has a meaning by itself."\cite{courant1996mathematics} 

These mathematical points will be very familiar to most readers; but my impression from papers and discussions is that how much our current physical theories say implicitly about time is not always adequately taken into account. \emph{My aim is to make explicit how our basic mathematics amounts to a theory of time}: a precondition for a clear discussion of whether any new physics of time is needed. Specifically, what we have just seen is that the representation of time as a real, independent variable -- something so familiar as to almost be unnoticed -- omits any thicker, dynamic notion of passage. As Courrant and Robbins put it, the $(\epsilon,\delta)$-definition ``leaves out something real to the intuition [of `a ``dynamic" notion of approach', but is] an adequate mathematical framework for expressing our knowledge of these concepts."\cite{courant1996mathematics}$^b$

Thus Cauchy's definition shows that the reals consistently represent the way in which a continuous function reaches a given value at a given time, and so (to that extent) consistently represent change. Again, our existing theories therefore adequately represent the physics of passage in this sense, without requiring any (imagined) thicker, dynamical notion. But the lesson of this analysis is even stronger, for it reveals how the thin notion of passage is in the very marrow of mathematical physics. It isn't simply that no thicker notion of passage is needed, it's that it would be difficult or impossible to provide one without abandoning the standard representation of time. At least, in the spirit of this section, introducing a dynamical notion of passage into physics would mean giving $t\to t_{now}$ some independent meaning and hence would require a new formal account of how functions obtain their limits. Whether such a new understanding (if even possible) was incompatible with Cauchy's definition and replaced it, or was compatible with it and expanded our representation, the result would be to put mathematical physics on a fundamentally different footing from that of the past 200 years. (Adopting discrete time won't by itself allow such a thicker, dynamical notion of passage. As I noted earlier, the question of reaching a limit, which motivated the attempt to thicken in the first place, does not even arise for a discrete series.) 

We have discussed continuous change, and the question of how a function reaches a limit, but similar points hold for other related features of change: for example, what is instantaneous velocity? Acceleration? How can motion start or end smoothly? All were historically important, and all found their answers in the nineteenth century articulation of the calculus, and hence all rest on the same structure of $\R$ exploited in the limit definition (and of course on the limit definition itself). Almost everything that we take for granted in formulating physical theories is at stake in a thickening of passage. When they acknowledge this situation, those searching for a physics of passage are faced with a challenge: either bring some suitable new mathematics to the table, or articulate a distinct sense of passage that preserves the mathematical foundations that we currently have. 

Perhaps most extant proposals are in the latter category: in particular, I think that causal set theory is best thought of this way. But then the exercise of being more precise about what one means by `passage' is crucial for developing such theories -- what, distinct from the notion discussed above is it? (Moreover, when philosophers invoke `passage' in a thick sense, they generally have in mind something that would require giving an independent meaning to $t\to t_{now}$, if it were to be given a mathematical form.) Of course one could simply say that passage cannot be fully captured mathematically; but our question is whether a mathematical physics of passage is needed, so that would be to say that we don't need -- because we can't have -- a physics of passage.

However, note (i) that Maudlin \cite{maudlin2007metaphysics} articulates the view that `movement or change or flow' of time escapes mathematical representation; but makes the case that such aspects of passage are nonetheless used in physics, specifically in explanation. A corollary would be that we cannot settle whether physics is compatible with passage by focussing on its mathematical representation, as I do here. Even more interestingly, given the argument of this section, (ii) Maudlin \cite{Mau:14} develops a new mathematical framework for time; perhaps it provides a more complete representation of passage. Lack of space prevents me from further discussion of this work, but it is of first importance for the question of a physics of passage.

\section{Extending the Minimal Model}\label{EMMT}

To review, the minimal model represents time by $\R$, which represents there being multiple times, and a differential structure that allows the $(\epsilon,\delta)$-definition of the limit. Now, the minimal model isn't really an adequate representation of time according to standard physical theories. First, time has metrical properties, at least in the sense that (in most theories) between two moments a measurable interval exists (whether it is intrinsic to time or `conventional' is not relevant to us); so our model of time must assign $\R$ a metric representing duration (for instance the natural one in which equal intervals of the reals represent equal intervals of time). Second, in relativistic theories there is no such thing as a single time that might be represented by $\R$, instead we have $\R^4$, with a timelike/spacelike distinction. But of course in this case it is timelike worldlines (or spacelike foliations) whose temporality is represented by the reals. And again, the standard model of particle physics involves a temporal asymmetry, in which case an orientation must also be added to the mathematical representation of time.$^c$

With these additional structures, the minimal model becomes the extended model, which represents all the usual features of time in physical theory. The majority view among philosophers -- including me -- is that none of these structures captures the (alleged) thicker, dynamical sense of passage that Price referred to. Many, though perhaps not a majority, also hold that we do not need any new physics of passage, because all its real features are already incorporated into the extended model, hence into existing physics. In this section I will discuss two arguments for the opposing view, with an eye to the question of how their ideas might be incorporated into physics: I will argue that the prospects are bleak.\\

Those who argue for a thicker notion often look to McTaggart \cite{McT:08} for a model of passage: one gloss on the difference between his `A-series' and `B-series' is that in the former but not the latter, time passes. In terms helpful for our discussion, a B-series representation of time involves the reals, with relations of past of, present at and future to between instants: such a model includes a temporal orientation of course, but the relations themselves are either timeless, or unchanging, as you prefer. The relative present is described (non-relativistically, at least) by the axiom:

\begin{equation}
\label{ }
(t)(t')\ t\ \textrm{is the present at}\ t'\ \mathrm{iff}\ t=t'
\end{equation}
(we need not consider whether this axiom \emph{identifies} relative presentness with =). The A-series supplements this model with an absolute, monadic property of presentness, which -- it is alleged -- applies to different moments at different times: the present changes, so time passes. (We temporarily put aside relativity here, to consider whether the idea even makes conceptual sense.) 

But McTaggart's absolute present is really no use at all in an account of passage. It's easy to locate the problem: simply ask which moment(s) are present? There are three possibilities, none of which is satisfactory for the proponent of the A-series: (i) no time is the absolute present; (ii) many times are the present; or (iii) exactly one time is the present. (i) leads back to a B-series. (ii) Of course \emph{at different instants}, different times are the present; but that is a relative statement, not an absolute one. Saying that two distinct instants are the absolute present is to say that two moments are simultaneous and yet at different times, which is a contradiction. (iii) Of course we expect that only one instant will actually be the present, but in that case, where is the passage? Again, different instants will be the present at different times, but that is just to say that different instants will be the present \emph{relative} to different times. To simply single out one time and ascribe an absolute property to it does not introduce passage. So the absolute present is impotent. (Not all philosophers accept this conclusion, and much more has been said on both sides; but to my mind, this basic framework demonstrates the underlying logic of the whole debate, and indicates the reasoning of those who reject an absolute present adequately for our purposes.)

So McTaggart does not provide a useful clue for incorporating passage into physics. But there is an alternative to both an absolute present and a presentness relation between instants in spacetime: that there is a second time, not identical to our physical time, but related to it. Speaking rather abstractly, let $t_O$ represent our ordinary time of experience, of clocks, and of current physics; and let $t_T$ represent a second, more fundamental time appearing in a theory from which our time emerges as a higher level phenomenon.$^d$ Then there could be relations between the two sets of times:  it could, for instance, be that $t_O=\mathrm{Monday}$ is the present at $t_T=0$, but not at $t_T=1$. In such a case, the O-present changes \emph{relative to} the T-time, even though the relations between the O-times are unchanging. The picture of dual times may seem somewhat outr\'e, but as we shall see below, something along these lines has arisen in causal set theory.\\

It is sometimes suggested that our personal \emph{experience} of time shows that we need a physics of passage, that something is missing from any current theories.$^e$ While there are indeed unique aspects of temporal experience, I do not believe that they show any such thing. I find in introspection a felt difference between temporal and spatial separations, between temporal and spatial relations -- especially, the formal may be directed, or causal -- and between temporal and spatial variation. I assume that you do too, though those with brain lesions may not. But (assuming that physicalism is true) these differences only require new physics if they cannot be understood in the context of existing physical theories, and that claim can be resisted. There is a significant literature explicating how such perceptual differences can be traced to physical properties of the brain: for instance its entropic nature, or the presence of motion detectors that are sensitive to variation over time, but not over space \cite{Huggett2010-HUGEAE} \cite{Hoe:}. These accounts make no appeal to any sense of passage that is not already admitted by existing physics. Admittedly, the project is difficult -- it's part of the mind-body problem! -- and a work in progress, but until some convincing reason is given to expect it to fail, that such aspects of temporal perception can be explained without any (new) physics of passage is a reasonable working hypothesis.

Sometimes, however, it seems that proponents of passage claim that we have awareness of something more than these concrete aspects of temporal experience, that we perceive \emph{passage itself}, as it were: Norton \cite{Nor:10} is a useful example. My first response is that my honest report of careful introspection is that I simply cannot find any aspect of my experience that corresponds to an awareness of temporal passage itself. I believe that my experience is typical in this matter, but I am willing to entertain the idea that my introspection may be faulty, or that my perceptual equipment may be defective: is it likely after all that there are perceptions of physical, dynamical passage itself?

This line of thought from temporal experience to a new physics of passage is in fact extremely tenuous. On the one hand we have the huge realm of phenomena that can be explained without a physics of passage; on the other a single phenomenon -- introspective awareness of a certain type -- that allegedly cannot. It would be an incredible coincidence if the one piece of evidence pointing us to physics beyond QM and GTR were to lie so easily within our grasp: no difficult observations, no complex experiments are needed, just some careful introspection. Note in particular that we have no other evidence that the brain requires a physics of passage for it to be understood; as far as we know, all neurological and biochemical processes of the brain are conventional physical ones. So the argument seems to be that a theory of \emph{consciousness} is not possible within the physics we have now, but requires a new physics. Of course, that thought would help explain why the mind-body problem is difficult, and indeed a connection to the problems of physics has been suggested before (e.g., `Wigner's friend'). But I seriously question whether research into fundamental physics is likely to be furthered by adding the problem of consciousness to the domain of study. Moreover, what is proposed is very different the discovery of anomalies in previous physical theories: the Michelson-Morley null-result, measurement of the anomalous perihelion of Mercury, and the photo-electric effect, for instance, required probing ever more carefully into the domains of existing theories. The claim that in perception we find a blatant anomaly in fundamental physics does not follow that pattern at all, and while logically possible  should be treated with the utmost skepticism.

Then what is the alternative explanation of reports of experienced dynamical passage? Since the reported experiences are introspective, the question is contentious, but I make the following proposal. A standard response from those skeptical of dynamical passage is to argue that there is an apparent experience of passage, but that it is an \emph{illusion} (e.g., as recently argued by Paul\cite{Paul2010-PAUTE}): one has the percepts that one would have if there were passage, but in fact there is not, so the percepts do not have an object -- they are produced instead by something other than passage. But this view grants too much to passage: it is more accurate to say that the reports of experience of passage are based on a \emph{mistake}. 

Too illustrate the difference, in the waterfall illusion$^f$ one has a percept of motion, when there is none; but if mistake a large dog for a bear, I do not have a percept of a bear, I have a percept of a dog, but misidentify its object. Moreover, if I pay more attention I can discover my mistake. My proposal is that much the same occurs in reports of the experience of passage: there is no percept of passage at all, with or without its object. Instead what are experienced are the characteristic aspects of temporal perception -- duration, causal connection, change and motion -- and these are misidentified as percepts of passage. More careful analysis (including understanding the psychology of perception) can correct this mistake too, and the percepts can be correctly identified for what they are.$^g$ The problem with the illusion view is that it admits the existence of a percept that could have passage as its object, and then must explain what causes the percept instead; the mistake view does not even allow such a percept. And really it's no surprise at all that something as complex and subtle as temporal perception should lead to mistakes.

In short, introspection shows that temporal experience is indeed quite singular, but it would be a huge and implausible leap from that observation to any new physics of passage; especially as the experiences are plausibly explicable within standard physics, and hence within the extended model of time.

\section{Is Causal Set Theory a Physics of Passage?}\label{CST}

The final section of this paper takes the discussion in a somewhat new direction, by considering a concrete proposal for a physics of passage. The proposal involves discrete spacetime, so is immune to the discussion of limits; but it draws on the dual time picture of passage suggested above, and has been linked to experience, so it illustrates our earlier considerations. I don't attempt to settle whether we have here a physics of passage, but just to sketch what it would take.\\

The proposal is based on an interpretation of `causal set theory' (CST); my presentation of the theory follows Dowker \cite{Dowker2005} \cite{Dow:14}. In brief, \emph{formally}, a causal set (`causet') is a collection of nodes related by an asymmetric, transitive `causal' relation; \emph{physically}, regions of phenomenal, relativistic spacetime are -- if looked at closely -- causets of discrete points, where two points are related iff one is in the causal past of the other. If they are not looked at closely, then causets appear to be continuous, as supposed in theories of classical spacetimes, the `phenomena' in the present context. In this way not only are the nodes of a causet interpreted as discrete points, but the formal `causal' relation between nodes has a physical interpretation in terms of phenomenal, relativistic causality, justifying its name.

How does one obtain a metrical spacetime from a structure as sparse as a causet? CST appeals to the fact that the causal structure and spacetime volume function of a continuous, semi-Riemannian spacetime, $\mathcal{S}$, determines a unique metric for $\mathcal{S}$. A causet comes with a discrete causal structure, and simply counting nodes provides a discrete volume function, so a causet is close to providing the necessary ingredients for fixing a metric. Using some smart mathematics, it is conjectured that in favorable cases a causet is actually close enough to determine the metric of a spacetime in which it can -- formally speaking -- be embedded. Note that both the manifold and the metric are derived, effective entities, while the fundamental story is one of discrete nodes and their non-metrical relations.

At least that is the story for a given causet. What's interesting for our purposes is that spacetime causets can have a (stochastic) dynamics \cite{rideout1999classical} according to which they \emph{grow} with respect to a time parameter, say $T$: if $q$ is in the causal future of $p$ then, with respect to $T$, $q$ joins the causet \emph{after} $p$.  (More specifically, $T$ is discrete, and counts the births of points.) The result of an evolution in $T$ is a causet and hence, as above, a region of spacetime, something which comes with metrical notions, including temporal extension. Thus (as I understand the creators of CST) there are two times: $T$, the `external' time in which spacetime grows, and the `internal time' given by the metric of the phenomenal spacetime itself. With respect to $T$, the amount of internal time is increasing: perhaps at $T=1$ generation there are no timelike curves in spacetime longer than 1s in duration, but there are after $T=100$ generations (David Rideout tells me that $T\gg10^{172}$ is more accurate!). $T$ is a fundamental parameter, which arises when a causet dynamics is given, while internal time is merely derived; hence these are two different quantities.

The connection to dynamical passage should be obvious: the CST dynamics apparently provides a physics for a dual time in the way we envisioned at the end of our discussion of McTaggart. In those terms, $t_O$ is the time internal to the spacetime derived from the causet, and $t_T$ is $T$, the time in which the causet grows. Then there are two kinds of temporal relations involving $t_O$ times: first, relations to other $t_O$ times, such as Monday being before Tuesday -- these are the ordinary, unchanging B-series relations. But second, there are relations between $t_O$ and $T$ times, between times in the causet and the external time: call these `$T$-relations'. Potentially, $t_O=0$s might be the present at $T=1$, but $t_O=1$s the present at $T=100$. With respect to $T$, the present changes, time passes. So far so good for the dual time account of passage.

Except, while phenomenal time is relativistic, $T$ is absolute: if $p$ and $q$ are not causally related, then there is no phenomenal fact about which came first, and yet there is a fact about which came into existence first with respect to $T$. Now, it could be that experiments probing the causet structure would reveal an absolute time, so that local Lorentz invariance is only approximate. But the Rideout-Sorkin dynamics is relativistic after all: it is designed to be `generally covariant' in the sense that the probability of any given spacetime is independent of the order in which the points of the causet are created. Put another way, \emph{there is no way to determine $T$ beyond what follows from the causal order given by the effective, internal time}: anything more is, in the general sense, pure gauge. Concretely, if $p$ is in the absolute future of $q$, then it is physical that $p$ is in the future of $q$ with respect to $T$, since every covariant ordering has $T(p)<T(q)$, but $|T(p)-T(q)|$ is unphysical. If $p$ and $q$ are spacelike, then there is not even a fact about which comes first with respect to $T$; $T(p)<T(q)$ and $T(q)<T(p)$ are physically indistinguishable.$^h$

Without further discussion, let's follow conventional wisdom and reject such gauge-equivalent differences as unphysical, mere choices of a representation (like selecting axes in order to work in coordinates). So the question of whether CST provides a \emph{physics} of passage must admit $T$ only up to gauge invariance, so that there are no $T$-relations  beyond what follows from internal time relations. Two options remain:

\begin{itemize}

  \item First, the relation of (for instance) in-the-future-of with respect to $T$ just \emph{is} the relation of being in-the-future-of with respect to time internal to the effective spacetime. In this case, CST introduces no additional temporal relations at all, and the dual-time strategy for implementing passage cannot get off the ground. To work it would have to be, for instance, not only that Monday is relatively earlier than Tuesday (a B-series fact), but \emph{also} that it occurs earlier with respect to $T$ (in order to get change with respect to $T$). But according to this option, those are the same thing, .
  
  \item  A second response holds that the $T$-relations, whilst fixed by the internal relations, are \emph{distinct} physical relations. In this view, there are two sets of relations because they arise at different levels (and play different logical roles in the theory), and hold between different things: the $T$-relations are fundamental, and hold just between the discrete nodes of the causet; the relations internal to spacetime are merely effective, and hold between the continuous points of the effective manifold.  Of course the relations agree, but that is a consequence of the general covariance of the dynamics which produces the causet, and does not make them the same relations.$^i$ (Analogously, maybe there are only quarters in my pocket, but arguably that doesn't make being-in-my-pocket and being-a-quarter-in-my-pocket the same property.)
  
 In this case the dual time strategy seems to be in business again, because the $T$-relations are not identical with the internal time relations: suppose, for instance, that $p$ occurs in none of the causets compatible with $T=0$, but in all the causets compatible with $T=100$, then with respect to $T$ things have changed because $p$ has gone from future to past. (Even though, in a spacetime containing $p$, it is timelessly, with respect to the internal time, in the future of the earlier points.)
  
\end{itemize}

In the first option, CST is not a physics of passage; if the second can be fleshed out, CST is a physics of passage. I want to stress that I do not claim to have pointed out any new formal feature of the theory: Dowker makes exactly the same point when she says that general covariance implies that `the physical order in which [things] happen is a partial order, not a total order' \cite{Dowker2005} (p.11). (Still less are my remarks any reason to question CST as a research program: they only seek to clarify a certain interpretation.) My point here is that making use of her proposal -- at least following the dual time approach -- requires taking $T$-relations and phenomenal spacetime relations as distinct, but in agreement (an agreement explained by the general covariance of the dynamics). Otherwise, CST provides no more structure to time than relativity, which is widely regarded not to permit passage. 

I have here only pointed to how a theory of passage might be started given two times, and work remains to be done to see if it can be properly carried out. In an interesting paper, Callender and W\"uthrich \cite{CalWut:} describe some of the problems and prospects for this kind of formal project (especially with regard to general covariance). But I say again, such a theory won't amount to an advance on relativity, to a new physics of passage, unless it takes the second option and takes fundamental $T$ time and relativistic time as literally two different aspects of physical reality.

Finally, I made some skeptical remarks above about appealing to direct experience of time to motivate a physics of passage, and the problems are illustrated here. On the one hand, if only internal time is physical, and experienced passage requires an appeal to a non-physical time, then we are outside physics, in the realm of dualism. On the other, if there are two physical times, then the one that is required for passage arises at the most fundamental level of physics, and it is incredible to suppose it has anything to do with consciousness. Either way, introspection provides a dubious guide for physics. (Dowker \cite{Dow:14} offers a cautious opposition to such skepticism.)

\section{Conclusions}

I hope that I have explicated some of the important ways in which time, and especially passage already appear in our theories -- and how the very mathematical framework resists one kind of program for developing a richer physics for passage. I've also indicated how certain intuitive pulls towards passage can be resisted: there's nothing actually wrong with the existing mathematics. Finally, in a somewhat more positive vein, I have sketched a different approach to passage; but even here the demands of relativity at least constrain progress when one thinks carefully about what passage means.

%\bibliographystyle{plainnat}
%\bibliography{PassBiblio}

\ \\

\noindent ENDNOTES:\\

a. Even the presentist -- who holds that only the present is `real' -- should agree, though they might conclude that therefore there is no mathematical representation of passage.\\

b. Also, ``no mathematician need or should lose the suggestive intuitive feeling that [`tend to', `approach' and `$\to$'] express" [306]. I'm unclear what role this `feeling' is supposed to play for the mathematician: not logical, clearly. Moreover, I reject their suggestion that some reality is missed by the mathematics: yes there is a limit to what the mathematics represents, but it does not follow that anything lies beyond.\\

c. Note that the $(\epsilon,\delta)$-definition of the limit is independent of any particular (smooth) metric, so the minimal model did not assume one. A similar point holds for the order provided by the $\leq$-relation: the minimal model assumes no temporal direction.\\

d. We will consider a possible example of this situation below; for a general discussion of spacetime emergence see \cite{HugWut:13}.\\

e. I have heard Lee Smolin make such a suggestion: specifically at the Centre for Philosophy of Science, at the University of Pittsburgh, 4/11/08. (His recent \cite{smolin2013time} does not repeat the idea.) \cite{Dow:14} entertains the idea without fully endorsing it.\\

f. E.g., http://www.georgemather.com/MotionDemos/MAEQT.html.\\

g. This point addresses Norton's argument \cite{Nor:10} that passage isn't an illusion because one can't lose it by perceiving differently. I agree! But one can correct one's mistake by more careful introspection and understanding of one's experiences. I made the same claim in \cite[\S10.5]{Huggett2010-HUGEAE}, and \cite{Hoe:} can be understood along similar lines, although he has a somewhat different analysis of the mistake. Maria Balcells \cite{Bal:13} also rejects the illusion view, but argues that the true objects of temporal perception that I listed, plus others, but no `dynamical passing', \emph{constitute} passage. In that case there is no mistake in identifying them as passage.\\

h. $T(p)$ denotes the $T$-time at which $p$ is created.\\

i. In technical terms, even if the relations are co-extensive, it doesn't immediately follow that they are identical (except on a narrow view of properties). And unless nodes and continuous points can be strictly identified, they are not even co-extensive.\\

\end{document}